\documentclass[fleqn,10pt]{SelfArx} 

\definecolor{color1}{RGB}{0,0,90} 
\definecolor{color2}{RGB}{0,20,20} 

\usepackage{hyperref} 
\hypersetup{hidelinks,colorlinks,breaklinks=true,urlcolor=color2,citecolor=color1,linkcolor=color1,bookmarksopen=false,pdftitle={Title},pdfauthor={Author},urlcolor=blue}

\usepackage{graphicx}
\usepackage{natbib}
\usepackage{amsmath}
\usepackage{multirow}
\usepackage{subfigure}
\usepackage{eucal}

\newcommand{\n}[1]{\mathrm{#1}}

\JournalInfo{Published in Journal of Magnetism and Magnetic Materials, Vol. 416, 321–324,, 2016} 
\Archive{\href{http://dx.doi.org/10.1016/j.jmmm.2016.05.034}{DOI: dx.doi.org/10.1016/j.jmmm.2016.05.034}} 

\PaperTitle{The magnetic properties of the hollow cylindrical ideal remanence magnet} 

\Authors{R. Bj\o{}rk} 
\affiliation{\textit{Department of Energy Conversion and Storage, Technical University of Denmark - DTU, Frederiksborgvej 399, DK-4000 Roskilde, Denmark}} 
\affiliation{*\textbf{Corresponding author}: rabj@dtu.dk} 

\newcommand{\keywordname}{Keywords} 

\Abstract{We consider the magnetic properties of the hollow cylindrical ideal remanence magnet. This magnet is the cylindrical permanent magnet that generates a uniform field in the cylinder bore, using the least amount of magnetic energy to do so. The remanence distribution of this magnet is derived and the generated field is compared to that of a Halbach cylinder of equal dimensions. The ideal remanence magnet is shown in most cases to generate a significantly lower field than the equivalent Halbach cylinder, although the field is generated with higher efficiency. The most efficient Halbach cylinder is shown to generate a field exactly twice as large as the equivalent ideal remanence magnet.}

\begin{document}

\flushbottom 

\maketitle 


\thispagestyle{empty} 

\section{Introduction}
Generating a large uniform magnetic field in the most efficient way possible is of key interest, both for scientific and commercial applications. A strong uniform magnetic field is required in a large number of applications, among these nuclear magnetic resonance (NMR) equipment, accelerator magnets and magnetic refrigeration devices. Typically, the magnetic field must be generated inside a cylindrical bore, and the surrounding magnet is in this case usually also cylindrical.

Previous investigations of such a cylindrical system have focussed on generating as large as field as possible in the cylinder bore \cite{Bloch_1998,Kumada_2001b,Bjoerk_2011b}, with little regard to the efficiency of the magnet design. In an efficient magnet design the  magnetic energy in the magnets is utilized fully to create the desired magnetic field. The efficiency of a permanent magnet design, $M$, can be defined as \cite{Jensen_1996}
\begin{equation}\label{Eq.Mstar_definition}
M=\frac{\int_{V_\n{field}}||\mathbf{B}||^2dV}{\int_{V_\n{mag}}||\mathbf{B_\n{rem}}||^2dV},
\end{equation}
where $V_\n{field}$ is the volume of the region where the magnetic field is created, $V_\n{mag}$ is the volume of the magnets, $\mathbf{B}$ is the magnetic field and $\mathbf{B_\n{rem}}$ is the remanence of the magnets used. The maximum value of $M$ is 0.25 \cite{Jensen_1996}.

Interestingly, the remanence distribution of the ideal cylindrical magnet that generates a uniform field in the cylinder bore, i.e. the design with the highest possible $M$, has been derived by Jensen \& Abele (1996) \cite{Jensen_1996}, but the field generated by this ideal remanence magnet have not been considered in detail. Little is known about this magnet design, even though it has been proven to be the most efficient design possible. It has never been realized physically, nor has it been investigated numerically or analytically in detail. Furthermore, only the magnetic efficiency, and not the magnetic field itself generated by this ideal remanence magnet, has been compared with the design most commonly used to generate a uniform field in a cylinder bore, namely the Halbach cylinder design. In this work we calculate the magnetic field generated by the ideal remanence magnet, and compare it to the Halbach cylinder, in order to determine the most optimal way to generate a desired field using the least amount of magnet energy.

Some of the magnetic structures that are considered in the following have a varying remanence through the magnetic structure. While these may not be easy to realize in a practical sense, they nevertheless generate the desired magnetic field using the least magnetic energy possible, and are therefore interesting to investigate scientifically.

\section{The ideal remanence cylindrical magnet}
Any ideal remanence magnet must have an irrotational and solenoidal remanence distribution, $\nabla{}\times{}\mathbf{B_\n{rem}}=0$ and $\nabla{}\cdot{}\mathbf{B_\n{rem}}=0$, and have no net magnetic charge on the surface of the magnet \cite{Jensen_1996}. Utilizing these requirement Jensen \& Abele (1996) \cite{Jensen_1996} were able to calculate the most efficient cylindrical magnet design, that generates a uniform field in the cylinder bore in two dimensions. We will only consider two dimensional structures in the following, i.e. flux leakage through the ends of the cylinder are ignored. This is a valid approximation as long as the cylinder is much longer than the diameter of the cylinder bore.

\begin{figure*}[!t]
   \includegraphics[width=2\columnwidth]{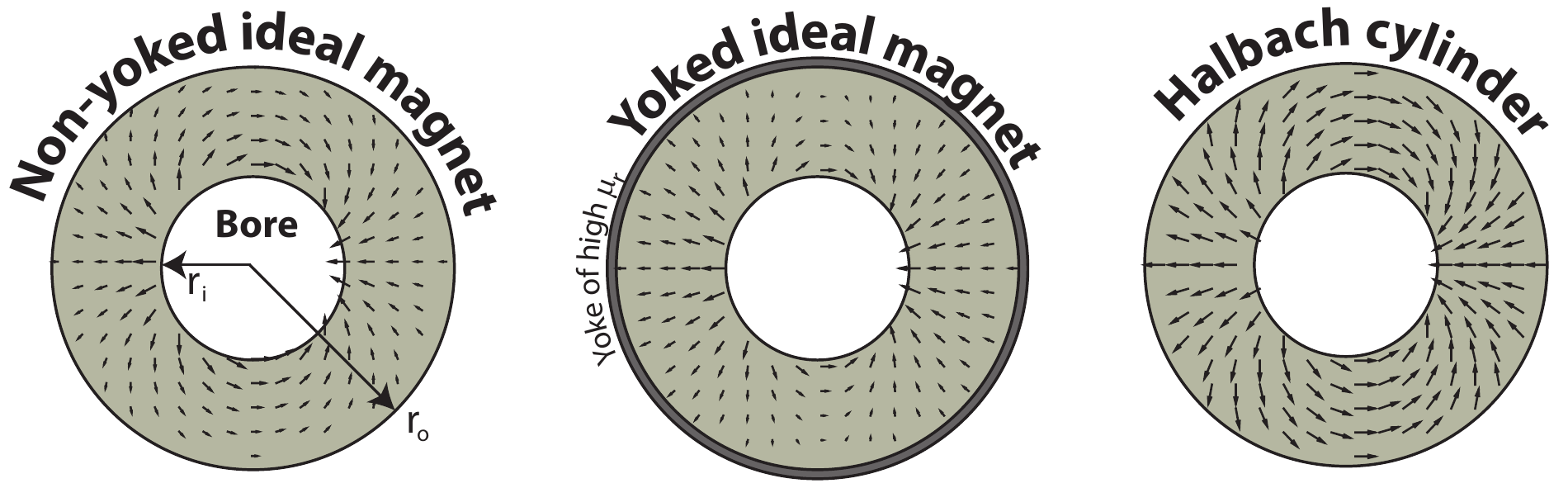}
      \caption{The remanence distribution of an ideal non-yoked and yoked magnet, as well as for a Halbach cylinder. The three magnets have the same maximum norm of the remanence.}
       \label{Fig.Drawing_all}
\end{figure*}

\subsection{Non-yoked design}
For the case where the cylindrical magnet is not surrounded by an iron yoke, the remanence potential for the ideal cylindrical magnet is given as \cite{Jensen_1996}
\begin{equation}
  \Phi_{J}^{*} = \frac{B_0r_\n{i}^2}{r_\n{o}^2-r_\n{i}^2}\left((\mu_r+1)\frac{r_\n{o}^2}{r}-(\mu_r-1)r\right)\cos(\phi)
\end{equation}
where $B_0$ is the norm of the generated flux density in the cylinder bore, $r_\n{i}$ and $r_\n{o}$ are the inner and outer radii of the magnet, respectively, $\mu_r$ is the relative permeability of the permanent magnet and $r$ and $\phi$ are polar coordinates. The remanence is given by
\begin{equation}\label{Eq.Phi_def}
  \mathbf{B_\n{rem}} = -\nabla{}\Phi_{J}^{*}
\end{equation}

Calculating the gradient in Eq. (\ref{Eq.Phi_def}), one get
\begin{eqnarray}
  B_\n{rem,r}     &= \mathcal{A}(r)\cos(\phi)\mathbf{\hat{r}}\nonumber\\
  B_\n{rem,\phi}  &= \mathcal{B}(r)\sin(\phi)\mathbf{\hat{\phi}}
\end{eqnarray}
where
\begin{eqnarray} \label{Eq.Coefficient_nonyoked}
  \mathcal{A}(r) &= -\frac{B_0r_\n{i}^2}{r_\n{o}^2-r_\n{i}^2}\left((\mu_r+1)\frac{r_\n{o}^2}{r^2}+(\mu_r-1)\right)\nonumber\\
  \mathcal{B}(r) &= -\frac{B_0r_\n{i}^2}{r_\n{o}^2-r_\n{i}^2}\left((\mu_r+1)\frac{r_\n{o}^2}{r^2}-(\mu_r-1)\right)
\end{eqnarray}
For the case of $\mu_r = 1$, the components $\mathcal{A}(r)$ and $\mathcal{B}(r)$ are identical. The remanence is seen to be a function of both $r$ and $\phi$. The remanence distribution is illustrated in Fig. \ref{Fig.Drawing_all}.

The norm of the remanence is given as
\begin{equation}
  ||\mathbf{B_\n{rem}}|| = \sqrt{\mathcal{A}(r)^2\cos^2(\phi) + \mathcal{B}(r)^2 \sin^2(\phi)}
\end{equation}
which simply reduces to $|\mathcal{A}(r)|$ for the case of $\mu_r = 1$.

Permanent magnets are limited in the maximum remanence that can be obtained. Therefore, it is of interest to determine the field that can be generated in the cylinder bore as function of the maximum remanence of the permanent magnet. As $\mu_r$ is always larger than or equal to 1 it can be seen from Eq. (\ref{Eq.Coefficient_nonyoked}) that $\mathcal{A}(r)\ge{}\mathcal{B}(r)$. This means that the maximum norm of the remanence will occur at $r=r_\n{i}$ and $\phi = [0,\; \pi]$. At these points, the norm is simply $|\mathcal{A}(r_\n{i})|$. The minimum norm of the remanence will always occur at $r=r_\n{o}$ and $\phi = [\pi/2,\; 3\pi/2]$. Here, the norm is simply $|\mathcal{B}(r_\n{o})|$.

Using that the maximum norm of the remanence is given by $|\mathcal{A}(r_\n{i})|$ and Eq. (\ref{Eq.Coefficient_nonyoked}), we get
\begin{equation}
  ||\mathbf{B_\n{rem,max}}|| = B_\n{rem,max} = \frac{B_0r_\n{i}^2}{r_\n{o}^2-r_\n{i}^2}\left((\mu_r+1)\frac{r_\n{o}^2}{r_\n{i}^2}+(\mu_r-1)\right)
\end{equation}
This is easily inverted in terms of $B_0$ as
\begin{equation}\label{Eq.B_0_nonyoked}
  B_0 = B_\n{rem,max}\frac{1-\frac{r_\n{i}^2}{r_\n{o}^2}}{\mu_r+1+\frac{r_\n{i}^2}{r_\n{o}^2}(\mu_r-1)}
\end{equation}
This is the magnitude of the uniform flux density generated by an ideal remanence magnet with a maximum remanence of $B_\n{rem,max}$.

For the case of $\mu_r = 1$, this reduces to
\begin{equation}
  B_0 = B_\n{rem,max}\frac{1-\frac{r_\n{i}^2}{r_\n{o}^2}}{2}
\end{equation}
while for an infinitely big magnet, $r_\n{i}/r_\n{o} \rightarrow 0$, so
\begin{equation}
  B_0 = B_\n{rem,max}\frac{1}{\mu_r+1}
\end{equation}
which for the case of $\mu_r = 1$ means that $B_0 = B_\n{rem,max}/2$. This is the maximum norm of the flux density that the ideal remanence magnet can generate for a given maximum remanence. The factor of 1/2 between the maximum remanence and the generated flux density is the maximum factor for a maximally efficient magnet \cite{Jensen_1996}.

\subsection{Yoked design}
The ideal distribution of remanence is different in the case that the cylindrical magnet is surrounded on the outside by a yoke of high permeability material. In this yoked case, the remanence potential given by \cite{Jensen_1996}
\begin{equation}
  \Phi_{J}^{*} = B_0r_\n{i}^2\left(\frac{\mu_r}{r_\n{o}^2-r_\n{i}^2}+\frac{1}{r_\n{o}^2+r_\n{i}^2}\right)\left(\frac{r_\n{o}^2}{r}-r \right)\cos(\phi)
\end{equation}

Calculating the remanence is similar to the non-yoked case, and the result is
\begin{eqnarray}
  B_\n{rem,r}     &= \mathcal{A}(r)\cos(\phi)\mathbf{\hat{r}}\nonumber\\
  B_\n{rem,\phi}  &= \mathcal{B}(r)\sin(\phi)\mathbf{\hat{\phi}}
\end{eqnarray}
where
\begin{eqnarray}
  \mathcal{A}(r) &= -B_0r_\n{i}^2\left(\frac{\mu_r}{r_\n{o}^2-r_\n{i}^2}+\frac{1}{r_\n{i}^2+r_\n{o}^2}\right)\left(\frac{r_\n{o}^2}{r^2}+1\right)\nonumber\\
  \mathcal{B}(r) &= -B_0r_\n{i}^2\left(\frac{\mu_r}{r_\n{o}^2-r_\n{i}^2}+\frac{1}{r_\n{i}^2+r_\n{o}^2}\right)\left(\frac{r_\n{o}^2}{r^2}-1\right)
\end{eqnarray}
The remanence distribution is illustrated in Fig. \ref{Fig.Drawing_all}. Again, the norm of the remanence will be largest at $r=r_\n{i}$ and $\phi = [0,\; \pi]$, where it will be $|\mathcal{A}(r_\n{i})|$. However, unlike the case for the non-yoked magnet, the components $\mathcal{A}(r)$ and $\mathcal{B}(r)$ are not identical for $\mu_r = 1$.

The equation for the maximum remanence becomes
\begin{equation}
  B_\n{rem,max} = -B_0\left(\frac{\mu_r}{r_\n{o}^2-r_\n{i}^2}+\frac{1}{r_\n{i}^2+r_\n{o}^2}\right)\left(r_\n{o}^2+r_\n{i}^2\right)
\end{equation}
which can be inverted in terms of $B_0$ as
\begin{equation}
   B_0 = B_\n{rem,max}\frac{1-\frac{r_\n{i}^2}{r_\n{o}^2}}{\mu_r+1+\frac{r_\n{i}^2}{r_\n{o}^2}(\mu_r-1)}
\end{equation}
This is exactly the same as the equation for the non-yoked case, i.e. Eq. (\ref{Eq.B_0_nonyoked}). Thus the non-yoked and the yoked cylindrical magnets generate the same magnetic field for the same choice of maximum remanence and the same size of the magnet. However, the magnetic energy in the permanent magnets is smaller in the yoked case, i.e. the yoked design have a higher efficiency.

\section{Comparing to the Halbach cylinder}
The Halbach cylinder is the most common way to generate a uniform magnetic field in a cylinder bore \cite{Mallinson_1973,Halbach_1980}. For this design, the remanence is given by
\begin{eqnarray}
B_\n{rem,r}    &= B_\n{rem}\cos(\phi)\mathbf{\hat{r}} \nonumber\\
B_\n{rem,\phi} &= B_\n{rem}\sin(\phi)\mathbf{\hat{\phi}},\label{Eq.Halbach_magnetization}
\end{eqnarray}
As can be seen, the norm of the remanence is uniform throughout the magnet, unlike the case of the ideal remanence magnets discussed above. The Halbach design has been used in a large number of applications including nuclear magnetic resonance (NMR) equipment \cite{Moresi_2003,Appelt_2006}, accelerator magnets \cite{Sullivan_1998,Lim_2005}, magnetic refrigeration devices \cite{Tura_2007,Bjoerk_2010b} and medical applications \cite{Sarwar_2012}.

For the Halbach cylinder, the magnetic flux density generated in the cylinder bore is given as \cite{Halbach_1980}
\begin{equation}\label{Eq.B_0_Halbach}
   B_0 = B_\n{rem}\ln{\left(\frac{r_\n{o}}{r_\n{i}}\right)}
\end{equation}

\begin{figure}[!t]
   \includegraphics[width=\columnwidth]{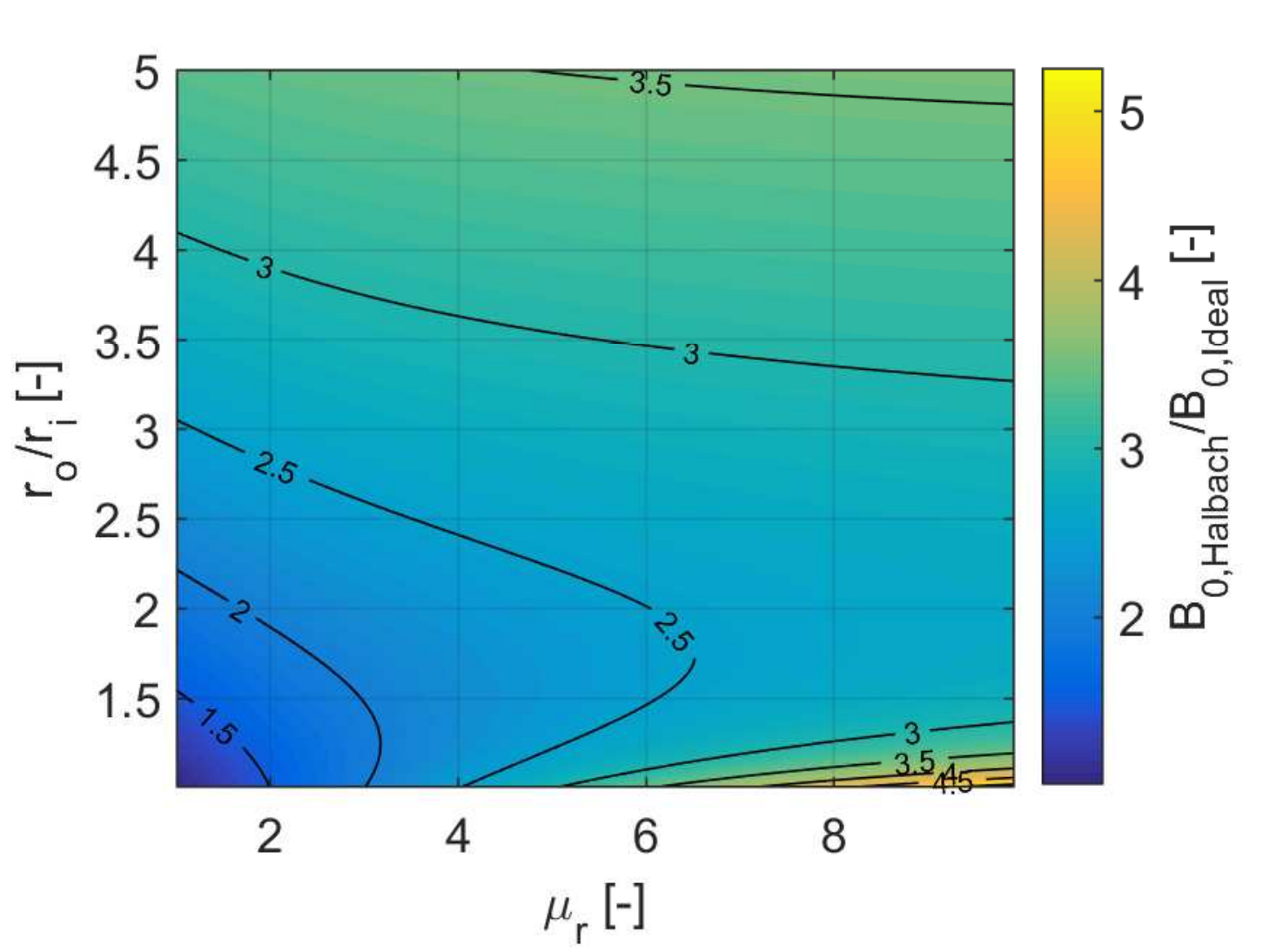}
      \caption{The ratio between the field generated by a Halbach cylinder and that generated by an ideal remanence magnet, as function of the size of the magnet and the relative permeability of the magnet material. The maximum remanence of the ideal remanence magnet is equal to the remanence throughout the Halbach magnet.}
       \label{Fig.B_ratio}
\end{figure}

The efficiency of the Halbach cylinder, as defined by $M$, is well known \cite{Bjoerk_2015a}, and the maximum value of the efficiency is $M\approx{}0.162$ for a ratio of the radii of $r_\n{o}/r_\n{i} \approx 2.2185$ \cite{Abele_1990,Coey_2003}. The efficiency of the ideal remanence magnets have previously been compared to that of the Halbach cylinder \cite{Jensen_1996}. Here it was shown that the ideal remanence magnets are always more efficient then the Halbach cylinder. However, the efficiency of the magnet design is of little interest if a magnetic field of the desired field strength cannot be generated in the cylinder bore. Therefore it is of critical importance to compare the actual magnitude of the magnet field generated in the cylinder bore in the different designs. Shown in Fig. \ref{Fig.B_ratio} is the ratio between the field generated by a Halbach cylinder and that generated by an ideal remanence magnet (both yoked and non-yoked as these generate the same field), for the case of the same maximum remanence and same size of the magnets. The ratio of the generated fields is shown as a function of the ratio of the inner and outer radius of the magnet as well as the relative permeability of the permanent magnet.

As can be seen from the figure, the Halbach cylinder generates a field that is always larger then that generated by the ideal remanence magnet. For this reason alone, the Halbach cylinder is preferential to the ideal remanence magnet, even though the latter has a higher efficiency, if the desired goal is to generate as high a field as possible.

It is of special interest to compare the two magnet design for the case of $\mu_r = 1$, as this is very close to the remanence of standard neodymium-iron-boron (NdFeB) magnets, where $\mu_r$ is in general taken to be 1.05 \cite{Standard}. Comparing the Halbach and ideal remanence magnets for $\mu_r = 1$ and assuming the maximum remanence of the ideal remanence magnet equal to the remanence throughout the Halbach magnet, we get
\begin{equation}
  \frac{B_{0,\n{Ideal}}}{B_{0,\n{Halbach}}} = \frac{1-\frac{r_\n{i}^2}{r_\n{o}^2}}{2\ln\left(\frac{r_\n{o}}{r_\n{i}}\right)}
\end{equation}

\begin{figure}[!t]
   \includegraphics[width=\columnwidth]{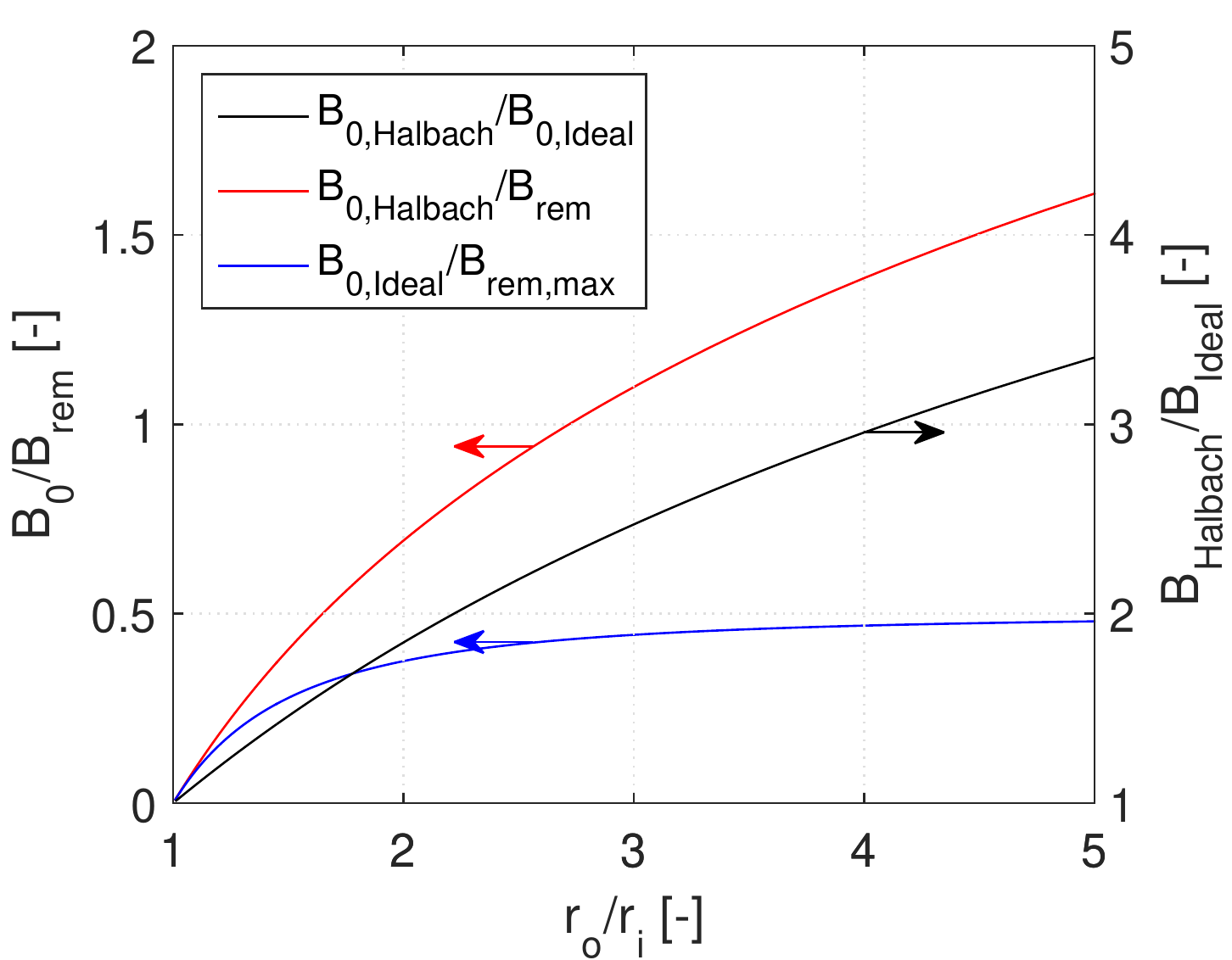}
      \caption{The ratio between the field generated by a Halbach cylinder and that generated by an ideal remanence magnet, as function of the size of the magnet for $\mu_r=1$. The fields generated, normalized by the remanence, are also shown. The maximum remanence of the ideal remanence magnet is equal to the remanence throughout the Halbach magnet.}
       \label{Fig.B_ratio_mur_1}
\end{figure}

This equation, along with Eq. (\ref{Eq.B_0_nonyoked}) and Eq. (\ref{Eq.B_0_Halbach}), for the ideal remanence magnet and the Halbach cylinder, respectively, are shown in Fig. \ref{Fig.B_ratio_mur_1} for $\mu_r = 1$. As can be seen from the figure, the Halbach cylinder always generate a substantially larger field then the ideal remanence magnet.

The ideal remanence magnet and the Halbach magnet can be compared in more detail for the size of the magnet where the Halbach cylinder is the most efficient. As per Bj\o{}rk et al. (2015) \cite{Bjoerk_2015a}, the optimal ratio of the radii for the most efficient Halbach is given as $(r_\n{i}/r_\n{o})_\n{opt} = e^{-W(-2e^{-2})/2-1}\approx{}0.4508$, where $W$ is the Lambert $W$-function. Using this expression in Eq. (\ref{Eq.B_0_nonyoked}) for the ideal remanence magnet, one can show that
\begin{equation}\label{Eq.B_0_at_roi_max_ideal}
  \left(\frac{B_{0,\n{Ideal}}}{B_\n{rem,max}}\right)_{\left[\frac{r_\n{o}}{r_\n{i}}\right]_\n{opt}} = \frac{1}{2}\left(1+\frac{W\left(-2e^{-2}\right)}{2}\right)
\end{equation}

For the case of the Halbach cylinder, Eq. (\ref{Eq.B_0_Halbach}), we get
\begin{equation}\label{Eq.B_0_at_roi_max_Halbach}
  \left(\frac{B_{0,\n{Halbach}}}{B_\n{rem}}\right)_{\left[\frac{r_\n{o}}{r_\n{i}}\right]_\n{opt}} = 1+\frac{W\left(-2e^{-2}\right)}{2}
\end{equation}

Comparing Eqs. (\ref{Eq.B_0_at_roi_max_ideal}) and (\ref{Eq.B_0_at_roi_max_Halbach}), one get
\begin{equation}
  \left(\frac{B_{0,\n{Ideal}}}{B_{0,\n{Halbach}}}\right)_{\left[\frac{r_\n{o}}{r_\n{i}}\right]_\n{opt}} = \frac{1}{2}
\end{equation}
This shows that at the optimal radius of the Halbach cylinder, i.e. the radius where it is the most efficient magnetically, the Halbach cylinder generates exactly twice the flux density generated by the ideal remanence magnet of the same size. Of course at this radius, the efficiency of the two designs are not identical. The efficiency of the Halbach is $M\approx{}0.162$, while it is $M\approx{}0.199$ for the non-yoked ideal remanence magnet and $M\approx{}0.240$ for the yoked ideal remanence magnet.

\section{Discussion and conclusion}
We have clearly shown that while the ideal remanence magnets are more magnetically efficient compared to the Halbach cylinder, they will always generate a lower field when compared with the equivalently sized Halbach cylinder. Thus the widespread use of the Halbach cylinder is justified, as the usual requirement in a application is to generate the largest field possible, and not use the magnetically most optimal design possible. Halbach cylinder are of course also easier to realize, as it has a constant remanence throughout the structure, making it more suitable for practical applications.

The inherent problem of the ideal remanence magnets is the low value of the generated field. By combining the ideal remanence magnet with a flux concentrating device, which is able to concentrate the field lines in a cylinder bore \cite{Navau_2012,Prat_2014}, the field generated by the ideal remanence magnet can be enhanced as desired. The flux concentrating device cannot change the efficiency of a given magnet design \cite{Bjoerk_2013a}, but will enhance the field generated by the permanent magnetic structure by a factor of $r_\n{o,con}/r_\n{i,con}$, where these are the outer and inner radii of the flux concentrator. Since the ideal remanence magnets can be made maximally efficiency (albeit only at infinite $r_\n{o}$), the desired field in the cylinder bore can be generated with maximum efficiency by fitting a flux concentrator of desired size in the cylinder bore of the magnet.

We have shown that the ideal remanence magnet that generates a uniform field in a cylinder bore will only generate a field substantially weaker than the maximum value of its remanence. Comparing with the Halbach cylinder, the optimum Halbach cylinder was shown to generate a field twice as large as the ideal remanence magnet of the same size.


\begin{thebibliography}{20}
\expandafter\ifx\csname natexlab\endcsname\relax\def\natexlab#1{#1}\fi

\bibitem{Bloch_1998}
F. Bloch, O. Cugat, G. Meunier and J. C. Toussaint, IEEE Trans. Magn. {\bf 34} (1998), 5.

\bibitem{Kumada_2001b}
M. Kumada, T. Fujisawa and Y. Hirao, Proc. Second Asian Part. Accel. Conf. (2001), 840.

\bibitem{Bjoerk_2011b}
R. Bj\o{}rk, J. Appl. Phys. {\bf 109} (2011), 013915.

\bibitem{Jensen_1996}
J. H. Jensen and M. G. Abele, J. Appl. Phys., {\bf 79} (1996), 1157.

\bibitem{Mallinson_1973}
J. C. Mallinson, IEEE Trans. Magn. {\bf 9 (4)} (1973), 678.

\bibitem{Halbach_1980}
K. Halbach, Nucl. Instrum. Methods, {\bf 169} (1980).

\bibitem{Moresi_2003}
G. Moresi and R. Magin, Concepts in Magn. Reson. Part B (Magn. Reson. Eng.), {\bf 19B} (2003), 35.

\bibitem{Appelt_2006}
S. Appelt, H. K\"uhn, F. W H\"asing, and B. Bl\"umich, Nat. Phys. {\bf 2} (2006), 105.

\bibitem{Sullivan_1998}
M. Sullivan, G. Bowden, S. Ecklund, D. Jensen, M. Nordby, A. Ringwall, and Z. Wolf, IEEE {\bf 3} (1998), 3330.

\bibitem{Lim_2005}
J. K. Lim, P. Frigola, G. Travish, J. B. Rosenzweig, S. G. Anderson, W. J. Brown, J. S. Jacob, C. L. Robbins, and A. M. Tremaine, Phys. Rev. Spec. Top. - Accel. Beams, {\bf 8} (2005), 072401.

\bibitem{Tura_2007}
A. Tura and A. Rowe, Proc. 2$^\mathrm{nd}$ Int. Conf. on Magn. Refrig. at Room Temp. (2007), 363.

\bibitem{Bjoerk_2010b}
R. Bj\o{}rk, C. R. H. Bahl, A. Smith, and N. Pryds, Int. J. Refrig. {\bf 33} (2010), 437.

\bibitem{Sarwar_2012}
A. Sarwar, A. Nemirovski, and B. Shapiro, J. Magn. Magn. Mater. {\bf 324 (5)} (2012), 742-754.

\bibitem{Bjoerk_2015a}
R. Bj{\o}rk, A. Smith and C. R. H. Bahl, J. Magn. Magn. Mater. {\bf 384} (2015), 128-132.

\bibitem{Abele_1990}
M. G. Abele and H. Rusinek, J. Appl. Phys., {\bf 67} (1990), 4644.

\bibitem{Coey_2003}
J. M. D. Coey and T. R. {Ni Mhiochain}, High Magnetic Fields (Permanent magnets), Chap. 2, p. 25, World Scientific (2003).

\bibitem{Standard}
Standard specifications for permanent magnet materials, Int. Mag. Assoc., Chicago, USA, (2000).

\bibitem{Navau_2012}
C. Navau, J. Prat-Camps, and A. Sanchez, Phys. Rev. Lett. {\bf 109 (26)} (2012), 263903.

\bibitem{Prat_2014}
J. Prat-Camps, C. Navau, and A. Sanchez, Appl. Phys. Lett. {\bf 105 (23)} (2014), 234101.

\bibitem{Bjoerk_2013a}
R. Bj{\o}rk, A. Smith, C. R. H. Bahl, J. Appl. Phys. {\bf 114 (5)} (2013), 053912.
\end{thebibliography}
\end{document}